\documentclass[twocolumn,showpacs,preprintnumbers,amsmath,amssymb,showkeys]{revtex4}

\usepackage{graphicx}
\usepackage{dcolumn}
\usepackage{bm}
\newcommand{\bqa}{\begin{eqnarray}}
\newcommand{\eqa}{\end{eqnarray}}
\newcommand{\be}{\begin{equation}}
\newcommand{\ee}{\end{equation}}

\begin{document}


\title{Decays of the X(3872) and $\chi_{c1}(2P)$ charmonium}

\author{Ce Meng$~^{(a)}$ and Kuang-Ta Chao$~^{(a,b)}$}
\affiliation{ {\footnotesize (a)~Department of Physics, Peking
University,
 Beijing 100871, People's Republic of China}\\
{\footnotesize (b)~China Center of Advanced Science and Technology
(World Laboratory), Beijing 100080, People's Republic of China}}


\begin{abstract}
We re-examine the re-scattering mechanism for the X(3872), as a
candidate for the 2P charmonium state $\chi_{c1}(2P)$, decaying to
$J/\psi\rho(\omega)$ through exchanging $D^{(*)}$ mesons between
intermediate states $D(\bar{D})$ and $\bar{D}^*(D^*)$. We evaluate
the dispersive part, as well as the absorptive one, of the
re-scattering amplitude and find that the contribution from the
dispersive part is dominant even when X(3872) lies above the
threshold of the neutral channel $th_n=m_{D^0}+m_{D^{*0}}$. We
predict $R_{\rho/\omega}\simeq 1$ for the $m_X$ region scanned by
experiments. Meanwhile, we also estimate the rate of $X\to
D^0\bar{D}^{0}\pi^0$. Our results favor a charmonium interpretation
of X(3872) when it lies slightly below the threshold of
$D^0\bar{D}^{*0}$. Furthermore, we evaluated the width of $X\to
J/\psi\rho$ with the help of a phenomenological effective coupling
constant $g_X$, and find the total width of X(3872) to be in the
range of 1-2~MeV.

\end{abstract}


\maketitle

\section{Introduction}
In recent years there have been a number of exciting discoveries of
new hadron states (for a recent review, see, e.g.~\cite{report}).
These discoveries are enriching and also challenging our knowledge
for the hadron spectroscopy, and the underlining theory for strong
interactions. Among these new states, the X(3872) may be the most
mysterious one, which was first discovered by the Belle
collaboration~\cite{belle03} in the invariant mass spectrum of
$J/\psi\pi^+\pi^-$ in the decay $B^+ \to J/\psi\pi^+\pi^-K^+$, and
confirmed soon by BaBar~\cite{BaBar05}, CDF~\cite{CDF04} and
D0~\cite{D004} collaborations. The world average mass is
$m_X=(3871.2\pm0.5)$ MeV and the width is $\Gamma_X<2.3$ MeV at
$90\%$ C.L.~\cite{PDG06}, which is consistent with the detector
resolution. The dipion mass distribution in $J/\psi\pi^+\pi^-$ seems
to favor a $\rho$ resonance for the dipion structure.  This implies
the $C$-parity of X(3872) is even, which is finally confirmed by the
measurement of $X(3872)\to \gamma
J/\psi$~\cite{belle05gammaJ,BaBar06gammaJ}. The angular distribution
analysis by Belle~\cite{belle05JPC} favors $J^{PC}=1^{++}$.
Analogous analysis~\cite{CDF06JPC} and the analysis for the dipion
mass spectrum~\cite{CDF05JPC} by CDF collaboration allow
$J^{PC}=1^{++}$ and $J^{PC}=2^{-+}$ as well.  The recent observation
of the near threshold decay $X\to
D^0\bar{D^0}\pi^0$~\cite{belle06DDpi} (with a slightly higher mass
of about 3875~MeV) by Belle may favor $J^{PC}=1^{++}$ but can not
rule out $J^{PC}=2^{-+}$. Moreover, Belle also see the sub-threshold
decay $X\to \omega J/\psi$ in  $B^+ \to J/\psi\pi^+\pi^-\pi^0
K^+$~\cite{belle05gammaJ}. So far, for the X(3872) four decay modes
have been observed with following
fractions\cite{PDG06,belle05gammaJ,belle06DDpi}
\begin{eqnarray}
{\cal B}(\!\!\!&B^\pm &\!\rightarrow K^\pm X) \times {\cal B}(X\to
\pi^{+}\pi^{-}J/\psi)   \nonumber \\   &=&(1.14\pm 0.20)\times
10^{-5}, \label{jpsipipi}
\end{eqnarray}
\be\label{psiomega}\frac{ {\cal B}(X\to J/\psi\pi^+\pi^-\pi^0)}{
{\cal B}(X\to J/\psi\pi^+\pi^-)}=1.0\pm 0.4\pm 0.3, \ee
\begin{eqnarray}
{\cal B}(\!\!\!&B&\rightarrow KX)\times{\cal B}(X\rightarrow D^0\bar
D^0\pi^{0})  \nonumber \\  &=&(1.22\pm0.31^{+0.23}_{-0.30})\times
10^{-4}, \label{DDpi}
\end{eqnarray}
\begin{equation}
 {\cal B}(X\rightarrow\gamma J/\psi)/ {\cal B} (X\rightarrow \pi^{+} \pi^{-}
J/\psi) = 0.14\pm 0.05, \label{Xrd}
\end{equation}
where the experimental value for $X\rightarrow\gamma J/\psi$ is
taken from the Belle measurement~\cite{belle05gammaJ}, while the
observed value of about 0.25 by BaBar is somewhat
larger~\cite{BaBar06gammaJ}.

For convenience, we define the following ratios and their values can
be deduced from (\ref{jpsipipi}), (\ref{psiomega}) and (\ref{DDpi}):
\be\label{Rpsiomega} R_{\rho/\omega}=\frac{\Gamma_
{\psi\rho}}{\Gamma_ {\psi\omega}}=1.0\pm 0.5, \ee
\be\label{RDDpi} R_{\rho/DD\pi}=\frac{\Gamma_ {\psi\rho}}{\Gamma_
{DD\pi}}=0.10\pm 0.05, \ee
where $\Gamma_{i}$ denotes the width of decay $X\to i$ with
$i=\psi\rho,~\psi\omega~\mbox{and}~D^0\bar{D}^0\pi^0$, respectively.

Because of the closeness of $m_X$ to the threshold
$M_{D^0\bar{D}^{*0}}=3871.81\pm 0.36$ MeV~\cite{CLEO07mD0}, many
authors identify the X(3872) with a molecule of
$D^0\bar{D}^{*0}+c.c.$ in S-wave~\cite{Tornqvist}, a loosely bound
state of charmed mesons. This is certainly a very attractive
interpretation, which also gives a natural explanation of the
$J^{PC}$ of $X(3872)$, and predicts $R_{\rho/\omega}\approx 1$ (see
Ref.~\cite{Swanson}) as well. Thus, the molecule becomes the most
popular interpretation for the X(3872). However, it seems to be
difficult for the molecule models to account for the large
production rates of X(3872) at B-factories and the
Tevatron~\cite{Bauer} unless ${\cal B}(X\to J/\psi\rho)$ is
large~\cite{Braaten}, which, however, seems to be in contradiction
with (\ref{RDDpi}). Furthermore, the molecule model predicted the
decay into $J/\psi\rho$ to be much superior to that into
$D^0\bar{D^0}\pi^0$, but this seems not to be supported by the
experiment. Moreover, the molecule model predicted that the
production rate of X(3872) in $B^+\to XK^+$ is much larger than that
in $B^0\to XK^0$~\cite{Braaten}, but the Belle data show that the
rate of $B^0\to XK^0$ with $X\to D^0\bar{D^0}\pi^0$ is approximately
equal to that of $B^+\to XK^+$ though the errors for the
measurements are large~\cite{belle06DDpi}. So, it might be useful to
try other possible interpretations for the X(3872).

Motivated by the large production rates in $B$ decays and in $p\bar
p$ collisions at the Tevatron, we suggested that the $X(3872)$ be a
$J^{PC} = 1^{++}$ $\chi_{c1}(2P)$ charmonium-dominated
state~\cite{Meng0506222}. This possibility has also been suggested
by Suzuki with detailed discussions on its decay
properties~\cite{suzuki}.
In this charmonium picture, the large rate of $B\to X(3872)K$, which
is comparable to (not much less than) $B\to \chi_{c1}(1P)K$, and the
similarity in production between $X(3872)$ and $\psi(2S)$ observed
by the CDF and D0 collaborations at the Tevatron can be well
understood~\cite{Meng0506222}. Leaving the mass
problem~\cite{Barnes,Eichten,CLQCD07,Li07} alone, the most difficult
problem of this assignment is how to explain the observed large
isospin violating effect expressed by $R_{\rho/\omega}\approx 1$,
since the state $\chi_{c1}(2P)$ is an isospin scalar.
Suzuki~\cite{suzuki} estimates that $R_{\rho/\omega}\approx 1/2$ in
an semi-quantitative way in which both $J/\psi\rho$ and
$J/\psi\omega$ are produced through the $D(D^*)$ exchange between
$D\bar{D}^*$ pair, and the large isospin violation can be accounted
for by the mass difference between neutral and charged $D\bar{D}^*$
thresholds and the large difference between the phase spaces of $X
\to J/\psi\rho$ and $X\to J/\psi\omega$. Recently, the ratio
$R_{\rho/DD\pi}$ was studied in a similar but more quantitative
way~\cite{Liu07}, and was predicted to be $R_{\rho/DD\pi}\approx
10^{-6}\mbox{-}10^{-4}$, which is far smaller than the experimental
data in (\ref{RDDpi}).

The estimation of $R_{\rho/DD\pi}$ given in Ref.~\cite{Liu07} is
based only on the imaginary part of the amplitude $\mathcal{A}(X\to
D^0\bar{D}^{*0}+c.c.\to J/\psi\rho)$. The quantity
$|\mbox{Im}\mathcal{A}|^2$ can be understood as the probability of
finding the final state $J/\psi\rho$ through re-scattering of the
$real$ $D^0\bar{D}^{*0}+c.c.$ pair in per unit of final-state phase
space. Then $\mbox{Im}\mathcal{A}$ is proportional to the phase
space factor of $X\to D^0\bar{D}^{*0}+c.c.$, which is small or even
zero, since the mass of $X$ is taken to be very close but above the
$D^0\bar{D}^{*0}$ threshold. As a consequence, the value of
$R_{\rho/DD\pi}$ given in Ref.~\cite{Liu07} is small. Furthermore,
since the charged channel $D^+\bar{D}^{*-}+c.c.$ is forbidden by
phase space, this mechanism will predict almost equal amplitudes for
$J/\psi\rho$ and $J/\psi\omega$, and then result in a large
$R_{\rho/\omega}$ of order 10 (or even larger) due to the difference
between  $J/\psi\rho$ and $J/\psi\omega$ phase spaces (see, e.g.
\cite{suzuki}).

On the other hand, in contrast to the imaginary part, the real part
of the re-scattering amplitude, which represents the effects of
virtual intermediate states $D\bar{D}^*$, may be dominant in this
case, since it does not suffer from the phase space suppression for
producing a real $D\bar{D}^*$ pair. Moreover, the real part may also
give a moderate value of $R_{\rho/\omega}$, since both the neutral
channel $D^0\bar{D}^{*0}+c.c.$ and the charged channel
$D^+\bar{D}^{*-}+c.c.$ can contribute through these virtual effects.

In this paper, we re-explore the re-scattering mechanism and
evaluate the real part of the amplitude as well as the imaginary
one. With a reasonable choice for the  phenomenological parameters,
we find that the experimental data in (\ref{Rpsiomega}) and
(\ref{RDDpi}) can be explained quite well if X(3872) is a 2P
charmonium state $\chi_{c1}(2P)$ lying below the threshold of
$D^0\bar{D}^{*0}$.

\section{The Model}
\subsection{$X\rightarrow J/\psi\rho(\omega)$}
\begin{figure}[t]
\begin{center}
\vspace{-2.5cm}
 \hspace*{-2.6cm}
\scalebox{0.5}{\includegraphics[width=28cm,height=35cm]{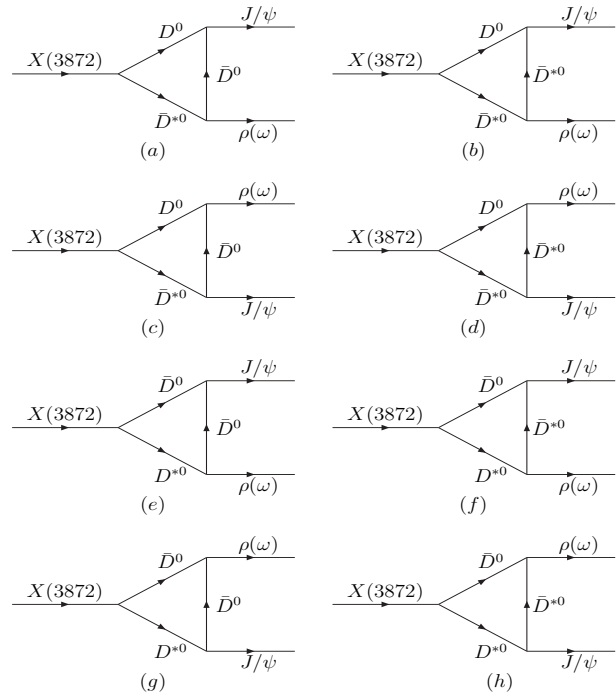}}
\end{center}
\vspace{-5.5cm}\caption{The decay diagrams for $X(3872)\to
\bar{D}^{*0}D^{0}+{\rm{c.c.}}\to
J/\psi\rho(\omega)$.}\label{Fig-psirho}
\end{figure}

In the re-scattering mechanism, the decay $X\to J/\psi\rho(\omega)$
can arise from exchange of a $D^{(*)}$ meson between $D(\bar{D})$
and $\bar{D}^*(D^*)$. The Feynman diagrams for $X\to
D^0\bar{D}^{*0}+c.c.\to J/\psi\rho(\omega)$ are shown in
Fig.~\ref{Fig-psirho}, and those involving charged intermediate
states can be easily obtained through replacements of
$D^0\bar{D}^{*0}$ by $D^+D^{*-}$ and $\bar{D}^0D^{*0}$ by
$D^-D^{*+}$ in Fig.~\ref{Fig-psirho}. Since $m_X$ is very close to
the threshold $M_{D\bar{D}^{*}}$ and the X(3872) couples to
$D\bar{D}^{*}$ in an S-wave, we assume that the re-scattering
contributions are dominated by the intermediate states
$D\bar{D}^{*}$, and those arising from higher exited $D$ meson
states are neglected.

Assume that X(3872) is the $\chi_{c1}(2P)$ state, and all the
vertexes in Fig.~\ref{Fig-psirho} are determined by the effective
Lagrangians, which are constructed based on the chiral and heavy
quark spin symmetries and parity conservation (for a review, see
Ref.~\cite{Casalbuoni}). These Lagrangians
read~\cite{Cheng05,Deandrea03} (for convenience here we use the same
notations and symbols as in Ref.~\cite{Liu07}):
\begin{subequations} \label{Lag1}
\begin{eqnarray}
\mathcal{L}_X&=&g_{X}X^{\mu}(DD^{*\dagger}_{\mu}-D^{\dagger}{D}^{*}_{\mu}),\label{LX}\\
\mathcal{L}_{\psi DD}&=&i g_{\psi DD} \psi_\mu \left(\partial^\mu
D{D}^{\dagger}-D\partial^\mu {D}^{\dagger}\right),\label{LpsiDD}\\
\mathcal{L}_{\psi D^*D}&=&\!\!-g_{\psi\!
D^*\!D}\varepsilon^{\mu\nu\alpha\beta}\partial_\mu \psi_\nu
\!\!\left(\partial_\alpha D^*_\beta {D}^{\dagger}\!\!
+ \!\! D \partial_\alpha {D}^{*\dagger}_\beta\!\! \right)\!,\label{LpsiD*D}\\
\mathcal{L}_{\psi D^*D^*}&=&-i g_{\psi D^* D^*} \Bigl\{ \psi^\mu
\left(\partial_\mu D^{*\nu} {D}_\nu^{*\dagger} -D^{*\nu}\partial_\mu {D}_\nu^{*\dagger} \right)\nonumber\\
&&+ \psi^\nu D^{*\mu}\partial_\mu{D}^{*\dagger}_{\nu} -
\psi_\nu\partial_\mu
D^{*\nu}  {D}^{*\mu\dagger}  \mbox{}   \Bigr\},\label{LpsiD*D*}\\
\mathcal{L}_{DDV}&=&-ig_{DDV}D_{i}^{\dagger}{\stackrel{\leftrightarrow}{\partial}}
_{\mu}D^{j}(\mathbb{V}^{\mu})^{i}_{j},\label{LDDV}\\
\mathcal{L}_{D^*DV}&=&-2f_{D^{*}DV}\varepsilon_{\mu\nu\alpha\beta}(\partial^{\mu}\mathbb{V}^{\nu})^{i}_{j}
(D_{i}^{\dagger}{\stackrel{\leftrightarrow}{\partial}}^{\alpha}D^{*\beta j}\nonumber\\
&&-D^{*\beta\dagger}_{i}{\stackrel{\leftrightarrow}{\partial}}^{\alpha}D^{j}),\label{LD*DV}\\
\mathcal{L}_{D^*D^*V}&=&+ig_
{D^{*}D^{*}V}D^{*\nu\dagger}_{i}{\stackrel{\leftrightarrow}{\partial}}_{\mu}
D^{*j}_{\nu}(\mathbb{V}^{\mu})^{i}_{j}\nonumber\\
&&+4if_{D^{*}D^{*}V}D^{*\dagger}_{i\mu}(\partial^{\mu}\mathbb{V}^{\nu}
-\partial^{\nu}\mathbb{V}^{\mu})^{i}_{j}D^{*j}_{\nu},\label{LD*D*V}
\end{eqnarray}
\end{subequations}
where the indexes $i,j$ in (\ref{LDDV}-\ref{LD*D*V}) represent the
flavors of light quarks, i.e. $D^{(*)}$=$(\bar{D}^{(*)0}$,
$D^{(*)-}$, $D_{s}^{(*)-})^T$, and they are hidden in
(\ref{LX}-\ref{LpsiD*D*}). $\mathbb{V}$ is
the $3\times 3$ matrix for the nonet vector meson, %
\begin{eqnarray}
\mathbb{V}&=&\left(\begin{array}{ccc}
\frac{\rho^{0}}{\sqrt{2}}+\frac{\omega}{\sqrt{2}}&\rho^{+}&K^{*+}\\
\rho^{-}&-\frac{\rho^{0}}{\sqrt{2}}+\frac{\omega}{\sqrt{2}}&
K^{*0}\\
K^{*-} &\bar{K}^{*0}&\phi
\end{array}\right).
\end{eqnarray}

We assume that chiral symmetry is preserved in (\ref{Lag1}). That
is, the coupling constants are blind to the flavor and there are no
isospin violations at the Lagrangian level. All the coupling
constants will be determined in the next section. However, it is
necessary to emphasize here that the determinations will not account
for the off-shell effect of the exchanged $D(D^*)$ meson, of which
the virtuality can not be ignored. As shown in Ref.~\cite{Liu07},
such effects can be accounted for by introducing, e.g., the monopole
form factors for off-shell vertexes. Let $q$ denote the momentum
transferred and $m_i$ the mass of exchanged meson, the form factor
can be written as~\cite{Liu07,Cheng05}
\begin{eqnarray}\label{formfactor}
\mathcal{F}(m_{i},q^2)=\bigg(\frac{\Lambda^{2}-m_{i}^2
}{\Lambda^{2}-q^{2}}\bigg),
\end{eqnarray}
and the cutoff $\Lambda$ can be parameterized as
\begin{eqnarray}\label{Lamvda}
\Lambda(m_{i})=m_{i}+\alpha \Lambda_{QCD}.
\end{eqnarray}

We are now in a position to compute the diagrams in
Fig.~\ref{Fig-psirho}.  If the X(3872) lies above the
$D^0\bar{D}^{*0}$ threshold, in the process $X(p_X,\epsilon_X)\to
D^{0}(p_{1})+\bar{D}^{*0}(p_{2},\epsilon_2)\to
J/\psi(p_{3},\epsilon_3)+\rho(\omega)(p_{4},\epsilon_4)$, where the
momenta $p$ and polarization vectors $\epsilon$ are denoted
explicitly for the mesons,  we can calculate the absorptive part
(imaginary part) of Fig.~\ref{Fig-psirho}(a) and find it to be given
by
\bqa\label{CutRule}
\textbf{Abs}(a_n)&=&\frac{|\vec{p}_1|}{32\pi^2m_X}\int d\Omega
\mathcal{A}(X\to D^0\bar{D}^{*0})\nonumber\\
&&\times \mathcal{A}_a(D^0\bar{D}^{*0}\to J/\psi\rho(\omega)), \eqa
where $\vec{p}_1$ is the 3-momentum of the on-shell $D^0$ meson in
the rest frame of X(3872) and the subindex "$n$" denotes the
contribution coming from the neutral channel (for the charged
channel we use "$c$"). Analogous expressions can be found for
diagrams (1b-1d), and the absorptive parts of diagrams (1e-1h) are
the same as those of diagrams (1a-1d), respectively. Explicitly, the
absorptive parts of diagrams (1a-1d) are given by
\begin{widetext}
\bqa\label{Abs}
\textbf{Abs}(a_n)&=&-\frac{|\vec{p}_1|}{32\pi^2m_X}\int
d\Omega(4\sqrt{2}g_X g_{\psi DD} f_{D^*
DV})\frac{\mathcal{F}^2(m_{1},q^2)}{q^2-m_1^2}(p_1\cdot\epsilon_3^*)\varepsilon_{\mu\nu\alpha\beta}p_4^\mu
\epsilon_4^{*\nu} p_2^\alpha \epsilon_X^\beta,\nonumber\\
\textbf{Abs}(b_n)&=&\frac{|\vec{p}_1|}{32\pi^2m_X}\int
d\Omega(\sqrt{2}g_X g_{\psi D^*D} g_{D^*
D^*V})\frac{\mathcal{F}^2(m_{2},q^2)}{q^2-m_2^2}\varepsilon_{\mu\nu\alpha\beta}p_3^\mu
\epsilon_3^{*\nu} p_1^\alpha \nonumber\\
&&\times\bigg\{(p_2\cdot\epsilon_4^*)\big[\epsilon_X^\beta-\frac{p_2\cdot\epsilon_X}{m_2^2}p_2^\beta\big]
+2r\big[p_4\cdot\epsilon_X-\frac{(p_2\cdot\epsilon_X)(p_2\cdot p_4)}{m_2^2}\big]\epsilon_4^{*\beta}
-2r\big[\epsilon_4^*\cdot\epsilon_X-\frac{(p_2\cdot\epsilon_X)(p_2\cdot \epsilon_4^*)}{m_2^2}\big]p_4^{\beta}\bigg\},\nonumber\\
\textbf{Abs}(c_n)&=&\frac{|\vec{p}_1|}{32\pi^2m_X}\int
d\Omega(\sqrt{2}g_X g_{\psi
D^*D}g_{DDV})\frac{\mathcal{F}^2(m_{1},q'^2)}{q'^2-m_1^2}
(p_1\cdot\epsilon_4^*)\varepsilon_{\mu\nu\alpha\beta}p_3^\mu\epsilon_3^{*\nu} p_2^\alpha \epsilon_X^\beta,\nonumber\\
\textbf{Abs}(d_n)&=&-\frac{|\vec{p}_1|}{32\pi^2m_X}\int
d\Omega(2\sqrt{2}g_X g_{\psi D^*D^*} f_{D^*
DV})\frac{\mathcal{F}^2(m_{2},q'^2)}{q'^2-m_2^2}\varepsilon_{\mu\nu\alpha\beta}p_4^\mu
\epsilon_4^{*\nu} p_1^\alpha \nonumber\\
&&\times\bigg\{2(p_2\cdot\epsilon_3^*)\epsilon_X^\beta
+\big[p_3\cdot\epsilon_X-\frac{(p_2\cdot\epsilon_X)(p_2\cdot
p_3)}{m_2^2}\big]\epsilon_3^{*\beta}
-\big[(\epsilon_3^*\cdot\epsilon_X)+\frac{(p_2\cdot\epsilon_X)(p_2\cdot
\epsilon_3^*)}{m_2^2}\big] p_3^{\beta}\bigg\}, \eqa
where the ratio $r=f_{D^*DV}/g_{D^*DV}$ and the momentum transferred
$q=p_3-p_1$, $q'=p_4-p_1$. The imaginary parts of the charged
channel amplitudes are the same as (\ref{Abs}) for $X\to
J/\psi\omega$ and of opposite signs for $X\to J/\psi\rho$. The total
absorptive part of the neutral (charged) channel amplitude can be
obtained by a simple summation and read
\bqa\label{Abs-total}
\textbf{Abs}_{n(c)}=2[\textbf{Abs}(a_{n(c)})+\textbf{Abs}(b_{n(c)})+\textbf{Abs}(c_{n(c)})+\textbf{Abs}(d_{n(c)})],\eqa
\end{widetext}
where the factor "2" comes from the equality of contributions from
diagrams (1a-1d) and (1e-1h).

The amplitudes in (\ref{Abs}) are almost equal to those given in
Ref.~\cite{Liu07} except that some minor errors in Ref.~\cite{Liu07}
have been corrected. All these amplitudes are proportional to the
phase space factor
\bqa\label{pCM}
\frac{|\vec{p}_1|}{m_X}&=&\frac{\sqrt{(m_X^2-(m_1+m_2)^2)(m_X^2-(m_1-m_2)^2)}}{2m_X^2}\nonumber\\
&&\simeq \sqrt{\frac{m_X-(m_1+m_2)}{2m_X}}\,, \eqa
which is very small even if X(3872) is above the $D^{0}\bar{D}^{*0}$
threshold.

In the case that X(3872) lies below the $D^{0}\bar{D}^{*0}$
threshold, the absorptive part (imaginary part) vanishes, and the
dispersive part (real part) of the re-scattering amplitudes will
play the role in the decay.

The dispersive part of the re-scattering amplitude can be obtained
from $\textbf{Abs}_{n}$ and $\textbf{Abs}_{c}$ via the dispersion
relation~\cite{suzuki,Cheng05}
\bqa\label{Dis}
\textbf{Dis}(m_X^2)\!=\!\frac{1}{\pi}\bigg(\!\!\int_{th_n^2}^{\infty}\!\!\frac{\textbf{Abs}_n(s')}{s'-m_X^2}ds'
\!\!+\!\!\int_{th_c^2}^{\infty}\!\!\frac{\textbf{Abs}_c(s')}{s'-m_X^2}ds'\bigg),\eqa
where $th_n=m_{D^0}+m_{D^{*0}}$ and $th_c=m_{D^\pm}+m_{D^{*\mp}}$
are the thresholds of neutral and charged channels respectively, and
the contributions arising from higher channels are neglected in
(\ref{Dis}) as we have mentioned before. Unlike the absorptive part,
the dispersive contribution suffers from the large uncertainties
arising from the complicated integrations in (\ref{Dis}). Since the
absorptive parts $\textbf{Abs}_{n(c)}(s)$ falls off as $s$
increases, it is reasonable to choose a cutoff for the integration
to make a numerical estimation. Following Ref.~\cite{suzuki}, we
choose the cutoff around $s_{max}=4m_{D^{*0}}^2$, which can shut the
widows of higher channels automatically.

It is worth emphasizing again that for $X\to J/\psi\omega$ the
contributions from neutral and charged channels are nearly equal and
share the same sign, while for $X\to J/\psi\rho$ they almost cancel
each other. This is not surprising since the explicit chiral
symmetry is maintained in the effective Lagrangians in (\ref{Lag1}).
So if we neglect the absorptive part, the isospin violation, which
is mainly due to the difference between $th_n$ and $th_c$ and that
between the thresholds of $J/\psi\rho$ and $J/\psi\omega$, seems to
be too small to account for the experimental data in
({\ref{Rpsiomega}}).
However, the large difference between the phase spaces of $X\to
J/\psi\rho$ and $X\to J/\psi\omega$ due to the large width of $\rho$
resonance may result in a favorable prediction for
$R_{\rho/\omega}$. To achieve this, we smear the width
$\Gamma_{\psi\rho(\omega)}^0(t)$, which is obtained through the
re-scattering amplitude in the narrow width approximation (NWA),
over the variable $t=m_4^2$ by the Breit-Wigner distribution as
\bqa \label{Breit-Wigner}
\Gamma(X&\to&\rho(\pi^+\pi^-)J/\psi)=\nonumber\\
&&\frac{1}{\pi}\int_{m_{2\pi}}^{(m_X-m_3)^2}\frac{\Gamma_{\psi\rho}^0(t)m_{\rho}\Gamma_{\rho}}
{(t-m_{\rho}^{2})^{2}+m_{\rho}^{2}\Gamma_{\rho}^{2}}dt,\nonumber\\
\Gamma(X&\to&\omega(\pi^+\pi^-\pi^0)J/\psi)=\nonumber\\
&&\frac{1}{\pi}\int_{m_{3\pi}}^{(m_X-m_3)^2}\frac{\Gamma_{\psi\omega}^0(t)m_{\omega}\Gamma_{\omega}}
{(t-m_{\omega}^{2})^{2}+m_{\omega}^{2}\Gamma_{\omega}^{2}}dt, \eqa
where $m_{2\pi}$ and $m_{3\pi}$ are the experimental cutoffs on the
$(\pi^+\pi^-)$ and the $(\pi^+\pi^-\pi^0)$ invariant masses
respectively, and $\Gamma_{\rho(\omega)}$ denotes the total width of
$\rho(\omega)$.

\subsection{$X\rightarrow D^0\bar{D}^0\pi^0$}
If X(3872) lies above the $D^0\bar{D}^{*0}$ threshold, the width of
$X\rightarrow D^0\bar{D}^0\pi^0$ can be given by
\be\label{Gamma-ddpi in NWA} \Gamma(X\!\!\to\!
D^0\!\bar{D}^0\!\pi^0)\!=\!2\Gamma(X\!\!\to\!
D^0\!\bar{D}^{*0})\mbox{Br}(\bar{D}^{*0}\!\!\to\! \bar{D}^0\pi^0),
\ee
where the branching ratio $\mbox{Br}(\bar{D}^{*0}\!\!\to\!
\bar{D}^0\pi^0)$ is known~\cite{PDG06} and the width
$\Gamma(X\!\to\! D^0\bar{D}^{*0})$ can be easily obtained from
$\mathcal{L}_X$ in the NWA:
\be\label{Gamma-ddstar in NWA} \Gamma(X\!\to\!
D^0\bar{D}^{*0})\!=\!\frac{g_X^2 |\vec{p}_1|}{24\pi
m_X^2}\big(3+\frac{|\vec{p}_1|^2}{m_{D^{*0}}^2}\big)\!\simeq\!\frac{g_X^2
|\vec{p}_1|}{8\pi m_X^2}, \ee
where the 3-momentum $\vec{p}_1$ is the same as in (\ref{CutRule}).
\begin{figure}[t]
\begin{center}
\vspace{-1.5cm}
 \hspace*{-1.5cm}
\scalebox{0.5}{\includegraphics[width=24cm,height=38cm]{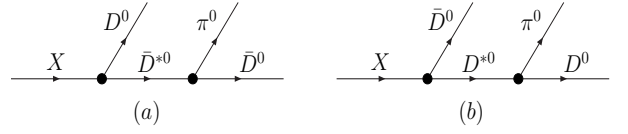}}
\end{center}
\vspace{-15.7cm}\caption{The diagrams for $X(3872)\to
D^{0}\bar{D}^{0}\pi$.}\label{Fig-DDpi^0}
\end{figure}

On the other hand, when $m_X$ is below the threshold $th_n$, it can
decay to $D^0\bar{D}^0\pi^0$ through virtual $D^{*0}(\bar{D}^{*0})$
as illustrated in Fig.~\ref{Fig-DDpi^0}~\cite{Colangelo07}. Here, we
need another effective Lagrangian to describe the
$D^0\bar{D}^{*0}\pi^0$ coupling~\cite{Casalbuoni,Colangelo07}:
\be\mathcal{L}_{D^*D\pi}=i\frac{g_{D^*D\pi}}{\sqrt{2}}\big(D^{*0}_\mu\partial^\mu
\pi^0\bar{D}^0-D^0\partial^\mu\pi^0\bar{D}^{*0}_\mu\big).\label{LD*Dpi}
\ee
Then the amplitude for $X(p_X,\epsilon_X)\!\to\!D^0
\bar{D}^0\pi^0(k_3)$ reads
\bqa\label{Amlitude-DDpi}
i\mathcal{M}&=&i(\mathcal{M}_a+\mathcal{M}_b) =\frac{i\sqrt{2}g_X
g_{D^*D\pi}}{q^2-m_{D^{*0}}^2+i
m_{D^{*0}}\Gamma(D^{*0})}\nonumber\\
&\times &\big[\frac{(q\cdot k_3)(q\cdot
\epsilon_X)}{m_{D^{*0}}^2}-(k_3\cdot\epsilon_X)\big], \eqa
where $q=p_X-p_1$ is the momentum transferred and $\Gamma(D^{*0})$
is the total width of $D^{{*0}}$. It can be verified that the
amplitude $\mathcal{M}$ generates the same width as that given in
(\ref{Gamma-ddpi in NWA}) in the limit $m_X-th_n\gg\Gamma(D^{*0})$.
However, the validity of (\ref{Amlitude-DDpi}) and
(\ref{Gamma-ddstar in NWA}) are questionable in the near threshold
region where $|m_X-th_n|\approx \Gamma(D^{*0})$ since the
perturbation calculations are known to be invalid in this region.

\section{Numerical Results and Discussions}
\subsection{Parameter determinations}

Since the numerical results are indeed sensitive to some of the
parameters introduced above, we need to explain how we determine
these parameters.

The coupling constants in Eq.~(\ref{LDDV}-\ref{LD*D*V}) are
universal for $\rho$ and $\omega$. They can be related to the
standard parameters in the so-called heavy meson chiral Lagrangian
\cite{Casalbuoni} through the relations \cite{Cheng05,Liu07}
\be\label{gDDV} g_{_{DDV}}=g_{_{D^{*}D^{*}V}}=\frac{\beta
g_{_{V}}}{\sqrt{2}},\;\;f_{_{D^{*}DV}}=\frac{f_{_{D^{*}D^{*}V}}}{m_{_{D^*}}}=\frac{\lambda
g_{_{V}}}{\sqrt{2}}.\nonumber\ee
The values of $g_V$, $\beta$, and $\lambda$ used here are the same
as in Ref.~\cite{Liu07} and are listed in Tab.~\ref{parameters}.
Similarly, the coupling constant in (\ref{LD*Dpi}) can be related to
the well-known parameter $g$ \cite{Casalbuoni,Colangelo07} through
the relation $g_{D^*D\pi}=2\sqrt{m_Dm_{D^*}}g/f_\pi$, where $f_\pi$
is the decay constant of $\pi$ and the value of $g$ can be
determined by the measurement of the width of $D^{*+}$
\cite{Colangelo07}. As a byproduct, we can estimate the total width
$\Gamma(D^{*0})\approx 0.07$ MeV by using the value of $g$ listed in
Tab.~\ref{parameters} together with the branching ratio
Br$(D^{*0}\to D^0\pi^0)$=$(61.9\pm 2.9)\%$~\cite{PDG06}.
\begin{figure}[t]
\begin{center}
\vspace{-1.2cm}
 \hspace*{-3.1cm}
\scalebox{0.5}{\includegraphics[width=28cm,height=35cm]{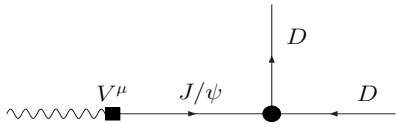}}
\end{center}
\vspace{-14.0cm}\caption{The diagram for calculating the matrix
element $\langle D|\bar{c}\gamma_{\mu}c|D\rangle$ in the VMD
mechanism.}\label{Fig-VMD}
\end{figure}

The coupling constant $g_{\psi DD}$ in (\ref{LpsiDD}) can be
estimated by the vector meson dominance (VMD) mechanism
\cite{Achasov94,Deandrea03}. Considering the matrix element $\langle
D|\bar{c}\gamma_{\mu}c|D\rangle$, it can be represented in
Fig.~\ref{Fig-VMD} in the VMD mechanism, where the circle vertex is
determined by Eq.~(\ref{LpsiDD}) and the box vertex is related to
the $J/\psi$ decay constant $f_{\psi}$ through the matrix element
$\langle
0|\bar{c}\gamma_{\mu}c|J/\psi(p,\epsilon)\rangle=f_{\psi}m_{\psi}\epsilon_\mu$.
Then at the normalization point, where the initial and final $D$
mesons have the same 4-velocities, we can determine $g_{\psi
DD}=m_\psi/f_\psi\simeq 8$, which is consistent with the prediction
of QCD sum rules \cite{Matheus02}. Other coupling constants in
(\ref{LpsiD*D}) and (\ref{LpsiD*D*}) can be estimated through heavy
quark symmetry relations: $g_{\psi D^*D^*}=m_D g_{\psi D^*D}=g_{\psi
DD}$~\cite{Deandrea03}. One should notice that  use of VMD here does
not mean that all higher resonances give contributions far smaller
than those from $J/\psi$, but it lies on the argument that these
contributions tend to cancel \cite{Deandrea03,Cea86}. For example,
the analogous effective coupling constant governing $\psi(3770)\to
D\bar{D}$ decay is about 3 times larger than $g_{\psi
DD}$~\cite{Colangelo07}.

The coupling $g_X$ is not involved in the ratios $R_{\rho/\omega}$
and $R_{\rho/DD\pi}$. However, $g_X$ is important for determining
the decay widths and clarifying the properties of X(3872). Assuming
that X(3872) is a pure charmonium 2P-state $\chi_{c1}(2P)$, we
parameterize $g_X=2\sqrt{2m_{D^0}m_{D^{*0}}m_X}g_1(2P)$, where
$g_1(2P)$ is the coupling constant governing the interactions of 2P
charmonium states with $D^{(*)}\bar{D}^{(*)}$~\cite{Colangelo07}. In
Ref.~\cite{Colangelo02-04}, the 1P partner of $g_1(2P)$ is estimated
in a similar way to that for $g_{\psi DD}$. One only needs to
replace the vector current $V^\mu=\bar{c}\gamma_{\mu}c$ by the
scalar one $S=\bar{c}c$, and the $J/\psi$ by the $\chi_{c0}$ in
Fig.~\ref{Fig-VMD}, and the result is \cite{Colangelo02-04}
\be\label{g1}
g_1(1P)=\sqrt{\frac{m_{\chi_{c0}}}{3}}\frac{1}{f_{\chi_{c0}}},
\ee
where the decay constant $f_{\chi_{c0}}$ is defined by $\langle
0|\bar{c}c|\chi_{c0}(p)\rangle=f_{\chi_{c0}}m_{\chi_{c0}}$. Using
$f_{\chi_{c0}}=(510\pm 40)$ MeV estimated by the sum rule
analysis~\cite{Colangelo02-04}, one can get $g_1(1P)\approx 2.1$
GeV${}^{-1/2}$. As we have mentioned, for the charm systems
$g_1(2P)$ should be of the same order as $g_1(1P)$. Then, the value
of $g_X$ can be estimated through $g_X\approx
2\sqrt{2m_{D^0}m_{D^{*0}}m_X}g_1(1P)\approx 23$ GeV. On the other
hand, the effective interactions between charmonium and
$D^{(*)}\bar{D}^{(*)}$ can also be estimated by the quark pair
creation models~\cite{yaouanc73}. From available calculations in
Refs.~\cite{Barnes,Eichten} together with Eq.~(\ref{Gamma-ddpi in
NWA}), we can deduce the effective coupling at hadronic level
$g_X\simeq$ 8-15 GeV when $\delta m_{X}=m_X-th_n$ varying  from 80
MeV to 0.5 MeV. Based on the two estimates mentioned above, we will
choose $g_X= 20$ GeV in our calculations. This should be a
reasonable choice for the coupling which describes the
$\chi_{c1}(2P)$ decay to $D\bar D^*$.

Since the virtuality of exchanged meson in Fig.~\ref{Fig-psirho} is
always larger than 1 GeV${}^2$, the amplitudes in (\ref{Abs}) are
sensitive to $\alpha$ when $\alpha<3$. The authors of
Ref.~\cite{Liu07} choose $\alpha$= 0.5-3.0. In
Ref.~\cite{Colangelo02-04}, it is argued that the value of $\Lambda$
in (\ref{formfactor}) can be around 3 GeV, which corresponds
$\alpha\approx 5$. In our calculations we choose $\alpha=4$.

For the charm meson masses we take $m_{D^0}=1864.847\pm 0.178$ MeV
\cite{CLEO07mD0} and  $m_{D^{*0}}-m_{D^0}=142.12\pm 0.07$
MeV~\cite{PDG06}, so the threshold $th_n$=3871.8MeV. For other mass
and width parameters, we refer them to PDG2006~\cite{PDG06}. The
cutoffs on dipion and tripion invariant masses in (\ref{Dis}) are
taken to be the same as in the Belle experiments
\cite{belle03,belle05gammaJ}:
\be\label{m2pi-m3pi}
m_{2\pi}=400\mbox{MeV},\;\;\;m_{3\pi}=750\mbox{MeV}. \ee
\begin{table}[t]
\caption{Parameters used in the calculations.}
\label{parameters}
\begin{tabular}{c|c|c|c|c}
\hline\hline $g_V$   &    $\beta$     &   $\lambda$   &  $g$    &
$g_{\psi DD}$    \\ \hline
    5.9    &    0.9   &  0.56 Gev$^{-1}$ &   0.6  &   8
    \\\hline\hline
       $g_{\psi D^*D}$   &  $g_{\psi D^*D*}$ &   $f_\pi$  &  $\Lambda_{QCD}$ &
$\alpha$   \\\hline
   4.3 Gev$^{-1}$   &   8  &   132 MeV  &  220 MeV  &  4 \\
\end{tabular}
\end{table}

\subsection{Numerical analysis}
\begin{figure}[t]
\begin{center}
\vspace{-1.0cm}
 \hspace*{-1.4cm}
\scalebox{0.5}{\includegraphics[width=22cm,height=32cm]{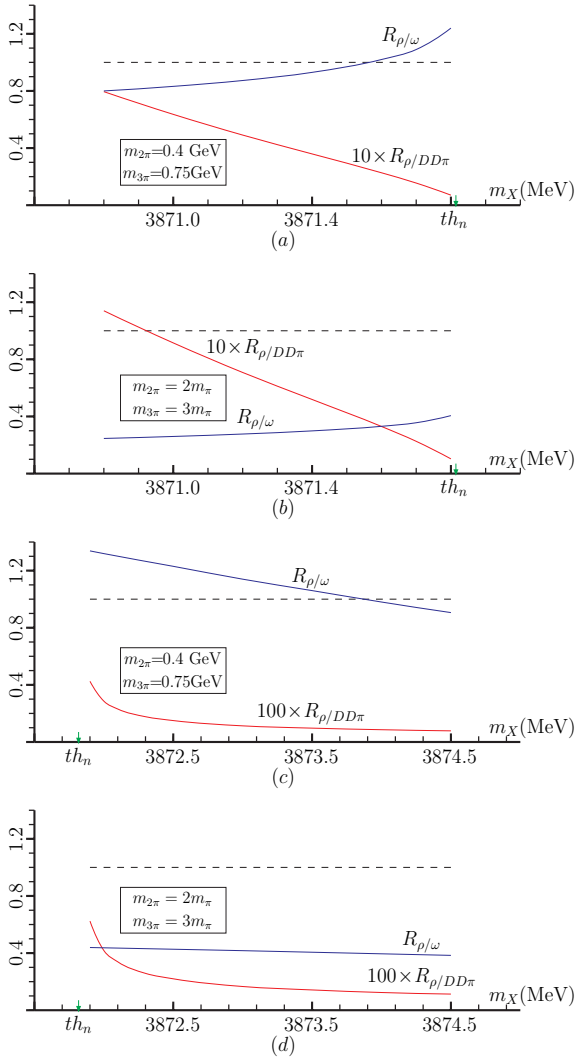}}
\end{center}
\vspace{-1.0cm}\caption{The $m_X$-dependence of $R_{\rho/\omega}$
and $R_{\rho/DD\pi}$. (a,b) for the region $m_X=$ 3870.8-3871.8 MeV
and (c,d) for $m_X=$ 3871.9-3874.5 MeV.}\label{Fig-R-mX}
\end{figure}

Our numerical results for the $m_X$-dependence of $R_{\rho/\omega}$
and $R_{\rho/DD\pi}$ are illustrated in Fig.~\ref{Fig-R-mX}. In the
below-threshold region where $m_X=$ 3870.8-3871.8 MeV,
Eq.~(\ref{Amlitude-DDpi}) is used to deduce the width
$\Gamma_{DD\pi}$. For the region above the threshold $th_n$ with
$m_X=$ 3871.9-3874.5 MeV, we use Eq.~(\ref{Gamma-ddpi in NWA}) and
(\ref{Gamma-ddstar in NWA}) to calculate $\Gamma_{DD\pi}$. The
contribution from absorptive part of the re-scattering amplitude is
not involved in Fig.~\ref{Fig-R-mX}, since it is numerically far
smaller than that from the dispersive part even after the
phase-space smearing in (\ref{Breit-Wigner}). For comparison, we
also choose the naive cutoffs $m_{2\pi}=2m_\pi$ and
$m_{3\pi}=3m_\pi$ to evaluate the integrations in
(\ref{Breit-Wigner}), and the results are shown in Fig.4(b) and
Fig.4(d). As usual, we use the central values of the parameters
given in the last subsection.

From Fig.~\ref{Fig-R-mX}(a,c), one can see that in both regions of
$m_X$,
\be\label{Prediction-Rrhoomiga} R_{\rho/\omega}=1.0\pm 0.3, \ee
which is consistent with Eq.~(\ref{Rpsiomega}). This result is
sensitive to the cutoff $m_{3\pi}$. For example, if we choose the
cutoffs given in Fig.~\ref{Fig-R-mX}(b,d), the width $\Gamma(X\to
J/\psi\omega)$ will be enlarged by a factor of 4, and the
corresponding value of $R_{\rho/\omega}$ is smaller than $0.4$.

Our prediction of $R_{\rho/\omega}$ in (\ref{Prediction-Rrhoomiga})
is a bit larger than that given in Ref.~\cite{suzuki}. It is not due
to a larger isospin violation in the dispersive part of the
re-scattering amplitude. In fact, here the isospin violation in the
dispersive part is only about $10\%$, which is smaller than that
expected in Ref.~\cite{suzuki} by a factor of 2. The difference is
mainly due to the fact that the momentum factors in vertexes $J/\psi
D^{(*)}D^{(*)}$ and $D^{(*)}D^{(*)}\rho(\omega)$ in
Fig.~\ref{Fig-psirho} are not considered in the semi-quantitative
estimation in Ref.~\cite{suzuki}. In fact, the re-scattering
$D\bar{D}^*\to J/\psi\rho(\omega)$ is a D-wave process, so that the
phase space smearing in (\ref{Breit-Wigner}) is more significant
than it is customarily expected.

Furthermore, one can see from Fig.~\ref{Fig-R-mX}(a) that the ratio
$R_{\rho/DD\pi}$ is roughly consistent with Eq.~(\ref{RDDpi}) except
for the very near-threshold region where $m_X=$ 3871.6-3871.8 MeV.
However, in this region, the width of $X\to D^0\bar{D}^0\pi^0$,
which is obtained from Eq.~(\ref{Amlitude-DDpi}), is questionable.
Roughly speaking, the charmonium picture of X(3872) is not in
serious contradiction with experimental data in (\ref{RDDpi}) if the
X(3872) is slightly below the $D^0\bar D^{*0}$ threshold, i.e.,
$m_X<th_n$.

\begin{figure}[t]
\begin{center}
\vspace{-2.2cm}
 \hspace*{-1.5cm}
\scalebox{0.5}{\includegraphics[width=22cm,height=35cm]{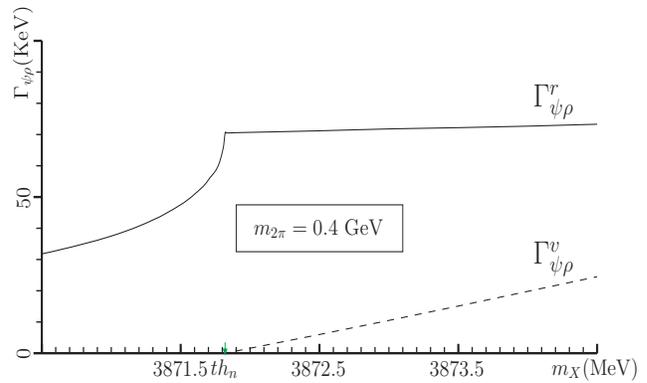}}
\end{center}
\vspace{-9.7cm}\caption{The width of $X\ \to J/\psi\rho$.
$\Gamma_{\psi\rho}^r$ arises from the real part of the re-scattering
amplitude and $\Gamma_{\psi\rho}^v$ from the imaginary part. Here
$g_X=20$ GeV is used.}\label{Fig-Gamma-psirho}
\end{figure}

In the region where $m_X>th_n$, the pure charmonium picture is
disfavored since the prediction of $R_{\rho/DD\pi}$ in
Fig.~\ref{Fig-R-mX}(c) is about two orders of magnitude smaller than
the experimental data in (\ref{RDDpi}). This is due to a rapid
increase of the decay rate of $X\to D^0\bar{D}^{*0}+c.c.$ as the
mass of X exceeds the $D^0\bar{D}^{*0}$ threshold.

We evaluate the width $\Gamma_{\psi\rho}$ by using $g_X=20$ GeV, and
the result is shown in Fig.~\ref{Fig-Gamma-psirho}.
One can see that $\Gamma_{\psi\rho}=$ 35-70 KeV. Then from the
obtained ratios in Fig.4 at the mass slightly below $th_n$ (say
around 3871.2~MeV) we have $\Gamma(X\to \psi\omega)=$ 25-100 KeV and
$\Gamma(X\to D^0\bar{D}^0\pi^0)=$ 250-1000 KeV. The width of $X\to
D^0\bar{D}^0\gamma$ can be estimated by a similar model shown in
Fig.~\ref{Fig-DDpi^0} and the value is no more than 300 KeV. Another
potential decay mode of the X(3872) is the inclusive light hadron
(LH) decay. We can use the available measurement of the width of 1P
state $\chi_{c1}(1P)$ \cite{PDG06} to roughly estimate that
$\Gamma(X\to \mbox{LHs})\simeq \Gamma(\chi_{c1}\to \mbox{LHs})\simeq
600$ KeV. The E1 transition width of $\chi_{c1}(2P)\to
\psi(2S)\gamma$ could be in the range 50-80~KeV; and the hadronic
transition width of $\chi_{c1}(2P)\to \chi_{c1}(1P)\pi\pi$ could be
10-100~KeV. Summing up all the estimated decay widths given above,
we find that the total width of X(3872) is about 1-2 MeV, which is
consistent with the experimental upper limit $\Gamma_X<2.3$ MeV.
Meanwhile, the branching ratio Br$(X\to J/\psi\rho)=(\mbox{2-7})\%$,
which can match the request of the large production rates at
B-factories and at the Tevatron~\cite{Meng0506222,suzuki,Bauer}.

Finally, it is worthwhile to mention that the E1 transition width
for $\chi_{c1}(2P) \to \gamma J/\psi$ can be estimated to be as
small as 7-11 KeV with relativistic corrections taken into
account\cite{Li07, Barnes}. Although this $2P-1S$ E1 transition rate
is sensitive to the model details duo to the node structure of
charmonium wavefunctions, there seems no difficulty in principle to
explain the ratio (\ref{Xrd}).


\section{Summary}
In summary, we re-examine the re-scattering mechanism for X(3872),
as a candidate for the 2P charmonium state $\chi_{c1}(2P)$, decaying
to $J/\psi\rho(\omega)$ through exchanging $D^{(*)}$ mesons between
intermediate states $D$ and $\bar{D}^*$ (or between $\bar{D}$ and
$D^*$). We evaluate the dispersive part, as well as the absorptive
one, of the re-scattering amplitude, and find that the contribution
from dispersive part is dominant even when X(3872) lies above the
threshold of the neutral channel $th_n=m_{D^0}+m_{D^{*0}}$. We
predict $R_{\rho/\omega}\simeq 1$ for the $m_X$ region scanned by
experiments. The prediction for $R_{\rho/DD\pi}$ favors $m_X<th_n$
and disfavors $m_X> th_n$, since in the latter case the prediction
is two orders of magnitude smaller than the experimental data due to
a much too large decay width into real $D^0\bar D^{*0}$ mesons.
Whereas when $m_X<th_n$ the $X$ can decay to $D^0\bar{D}^{0}\pi^0$
only through a virtual $\bar{D}^{*0}$ and a $D^0$, and therefore the
decay width of $X\to D^0\bar{D}^{0}\pi^0$ becomes much milder.
Furthermore, we evaluated the width of $X\to J/\psi\rho$ with the
help of a phenomenological effective coupling constant $g_X$, which
can be estimated from two different ways related to P-wave
charmonium decaying into two charmed mesons. We find that the total
width of the X(3872) is in the range of 1-2~MeV, and the theoretical
results for the four decay channels are roughly consistent with
experimental ratios (5) and (6), as well as (4). The remaining
problem is how to accept the low mass of X(3872) as the candidate of
a $\chi_{c1}(2P)$-dominated state. As shown in e.g. Table I of
Ref.\cite{Barnes}, the mass splitting between $\chi_{c1}(2P)$ and
$\chi_{c2}(2P)$ is predicted to be about 30~MeV in potential models
without the coupled channel effects. Including the mass shifts due
to coupled channel effects one finds that the mass of
$\chi_{c1}(2P)$ could be further lowered by about 30~MeV relative to
that of $\chi_{c2}(2P)$ and results in a mass difference between
$\chi_{c2}(2P)$ and $\chi_{c1}(2P)$ of about 60~MeV (detailed
discussions will be presented in Ref.\cite{Li07}). The Z(3930) meson
observed by Belle has been identified as the $\chi_{c2}(2P)$
charmonium \cite{bellechic2}, and if the above estimated mass
splitting makes sense, then the mass of $\chi_{c1}(2P)$ will be
around 3872~MeV. We will leave the mass issue to be discussed
elsewhere. Finally, based on the obtained results, we tend to
conclude that a $\chi_{c1}(2P)$-dominated state could be compatible
with the observed decays, production in the $B$ decay and at the
Fermilab Tevatron, and even the mass of the X(3872). Therefore,
aside from the molecule and other interpretations, the
$\chi_{c1}(2P)$ charmonium-dominated state could still be a possible
assignment for the X(3872).

\begin{acknowledgments}
We wish to thank B. Zhang, H.Q. Zheng, and S.L. Zhu for helpful
discussions. We also thank G. Bauer for a useful comment on the
experimental determination of the X(3872) quantum numbers. This work
was supported in part by the National Natural Science Foundation of
China (No 10421503, No 10675003), the Key Grant Project of Chinese
Ministry of Education (No 305001), and the Research Found for
Doctorial Program of Higher Education of China.
\end{acknowledgments}

\bibliography{apssamp}

\end{document}